\documentclass[%
aip,jap,
 amsmath,amssymb,floatfix,
 preprint,%
]{revtex4-1}
\usepackage{multirow}
\usepackage{amsmath,amssymb,amsfonts}
\usepackage{algorithmic}
\usepackage{array}
\usepackage{xcolor}
\usepackage{footnote}
\usepackage{threeparttable}
\usepackage{graphicx}
\usepackage{textcomp}
\usepackage{xcolor}
\usepackage{array}
\usepackage{mathtools}
\usepackage{siunitx}

\DeclarePairedDelimiter\abs{\lvert}{\rvert}%

\def\BibTeX{{\rm B\kern-.05em{\sc i\kern-.025em b}\kern-.08em
    T\kern-.1667em\lower.7ex\hbox{E}\kern-.125emX}}
\begin{document}

\title{Spin Wave Based Approximate $4$:$2$ Compressor
{}
}

\author{Abdulqader Mahmoud}
\email{A.N.N.Mahmoud@tudelft.nl}
\affiliation{Delft University of Technology, Department of Quantum and Computer Engineering, 2628 CD Delft, The Netherlands}

\author{Frederic Vanderveken}
\affiliation{KU Leuven, Department of Materials, SIEM, 3001 Leuven, Belgium}
\affiliation{Imec, 3001 Leuven, Belgium}

\author{Florin Ciubotaru}
\affiliation{Imec, 3001 Leuven, Belgium}

\author{Christoph Adelmann}
\affiliation{Imec, 3001 Leuven, Belgium}

\author{Said Hamdioui}
\affiliation{Delft University of Technology, Department of Quantum and Computer Engineering, 2628 CD Delft, The Netherlands}

\author{Sorin Cotofana}
\email{S.D.Cotofana@tudelft.nl}
\affiliation{Delft University of Technology, Department of Quantum and Computer Engineering, 2628 CD Delft, The Netherlands}

\begin{abstract}
In this paper, we propose an energy efficient SW based approximate $4$:$2$ compressor comprising a $3$-input and a $5$-input Majority gate. We validate our proposal by means of micromagnetic simulations, and assess and compare its performance with one of the state-of-the-art SW, \SI{45}{nm} CMOS, and Spin-CMOS counterparts. The evaluation results indicate that the proposed compressor consumes $31.5$\% less energy in comparison with its accurate SW design version. Furthermore, it has the same energy consumption and error rate as the approximate compressor with Directional Coupler (DC), but it exhibits $3$x lower delay. In addition, it consumes $14$\% less energy, while having $17$\% lower average error rate than the approximate \SI{45}{nm} CMOS counterpart. When compared with the other emerging technologies, the proposed compressor outperforms approximate Spin-CMOS based compressor by $3$ orders of magnitude in term of energy consumption while providing the same error rate. Finally, the proposed compressor requires the smallest chip real-estate measured in terms of devices
\end{abstract}

\maketitle

\section{Introduction}
The information technology revolution has led to a rapid raw data rapid increase, which processing calls for high performance computing platforms \cite{data1}. Up to date, downscaling Complementary Metal Oxide Semiconductor (CMOS) has been effective to satisfy these requirements, however, Moore’s law has reached its near economical end as CMOS feature size reduction is becoming increasingly difficult due to leakage, reliability, and cost walls \cite{cmosscaling1}. As a result, different technologies have been investigated to replace CMOS such as graphene devices \cite{Yande1}, memristor \cite{memristor10}, and spintronics \cite{ITRS}. In this paper, we chose to study one type of spintronics technology, the Spin Wave (SW) technology, which appears to open the way towards the most energy efficient digital computing paradigm \cite{amahmoud2,amahmoud1,parallelism,fanout10}. SW based computing is promising for three main reasons \cite{amahmoud2,amahmoud1,parallelism,fanout10}: 1) it has ultra-low energy consumption potential because it does not rely on electrons movements but just on their spinning around the magnetic field orientation \cite{amahmoud2,amahmoud1,parallelism,fanout10}, 2) it is highly scalable because SW's wavelength (which is the distance between two electrons that exhibit the same behavior) can reach the nanometer scale \cite{amahmoud2,amahmoud1,parallelism,fanout10}, and 3) it has an acceptable delay \cite{amahmoud2,amahmoud1,parallelism,fanout10}. As a consequence of these promising features, different researcher groups have made use of SW interaction to build logic gates and circuits.

The first experimental SW logic gate is an inverter, designed by utilizing a Mach-Zehnder interferometer \cite{logic21}. Moreover, the Mach-Zehnder interferometer has been used to build a single output Majority, (N)AND, (N)OR, and X(N)OR gates \cite{logic21}, while multi-output SW logic gates have been introduced in \cite{fanout, fanout10,fanout11}. Furthermore, multi-frequency logic gates that enhance SW computing and storage capabilities have been proposed in \cite{parallelism,parallelism1}, and wavepipelining has been achieved with pulse mode operation in the SW domain by utilizing four cascaded Majority gates \cite{wavepipeline}. In addition, different SW circuits have been also demonstrated at conceptual level  \cite{logic1}, simulation level \cite{mahmoud2021spin,amahmoud1}, and practical millimeter scale prototypes \cite{memory3}. All the aforementioned logic gates and circuits were designed to provide accurate results; however, many applications such as multimedia processing and social media are error-tolerant, and within certain error limits, they still function correctly \cite{applications}. Hence, those applications can benefit from approximate computing circuits, which save significant energy, delay, and area.

Based on the previous discussion on the SW technology potential and the approximate computing benefits one can conclude that SW approximate circuits are of great interest. In view of this observation, and given that multiplication is heavily utilized in error tolerant applications, and fast state-of-the-art multipliers are build with 4:2 compressors \cite{conventional_compressor4} we introduce in this paper a novel approximate SW $4$:$2$ compressor. The paper main contributions can be summarized as follows:
\begin{itemize}
  \item Developing and designing an approximate SW $4$:$2$ compressor: We propose an approximate $4$:$2$ compressor consisting of two Majority gates that provides an average error rate of $31$\%.
  \item Enabling directional couplers free approximate circuit design: We demonstrate that Majority gates can be directly cascaded, i.e., without amplitude normalization of domain conversion, to form a $4$:$2$ compressor with no additional average error rate penalty. 
  \item Validating the proposed $4$:$2$ Compressor: We demonstrate by means of MuMax3 micromagnetics simulations the correct functionality of the proposed approximate $4$:$2$ compressor.  
  \item Demonstrating the superiority: The proposed approximate SW $4$:$2$ Compressor performance is assessed and compared with state-of-the-art SW, \SI{45}{nm} CMOS, and Spin-CMOS counterparts. The evaluation results indicate that the proposed compressor saves $31.5$\% energy in comparison with the accurate SW design, whereas it has the same energy consumption and error rate as the approximate compressor with Directional Coupler (DC), but while being 3x faster. In addition, the proposed compressor consumes $14$\% less energy while providing $17$\% less error rate when compared with the approximate \SI{45}{nm} CMOS  counterpart. Moreover, the proposed compressor outperforms approximate Spin-CMOS equivalent design by $3$ orders of magnitude in terms of energy while having the same error rate. Finally, the proposed compressor requires the smallest chip real-estate.
\end{itemize}

The rest of the paper is organized as follows. Section \ref{sec:Basics of spin-wave technology} explains SW computing background. Section \ref{sec:Proposed approximate functions} introduces the proposed approximate $4$:$2$ compressor and Section \ref{sec:Simulation Setup and Results} provides inside on the simulation setup and results. Section \ref{sec:Performance Evaluation} reports performance evaluation and comparison with state-of-the-art data. Section \ref{sec:Conclusion} concludes the paper.

\section{Spin Wave Based Technology Fundamental and Computing Paradigm}
\label{sec:Basics of spin-wave technology}

The magnetization dynamics caused by the magnetic torque when the magnetic material magnetization is out of equilibrium is captured by the Landau-Lifshitz-Gilbert (LLG) equation \cite{amahmoud2}:

\begin{equation} \label{eq:1}
\frac{d\vec{M}}{dt} =-\abs{\gamma} \mu_0 \left (\vec{M} \times \vec{H}_{eff} \right ) + \frac{\alpha}{M_s} \left (\vec{M} \times \frac{d\vec{M}}{dt}\right ),
\end{equation}
where $\gamma$ is the gyromagnetic ratio, $\mu_0$ the vacuum permeability, $M$ the magnetization, $M_s$ the saturation magnetization, $\alpha$ the damping factor, and $H_{eff}$ the effective field, which consists of the external field, the exchange field, the demagnetizing field, and the magneto-crystalline field.

\begin{figure}[t]
\centering
  \includegraphics[width=0.75\linewidth]{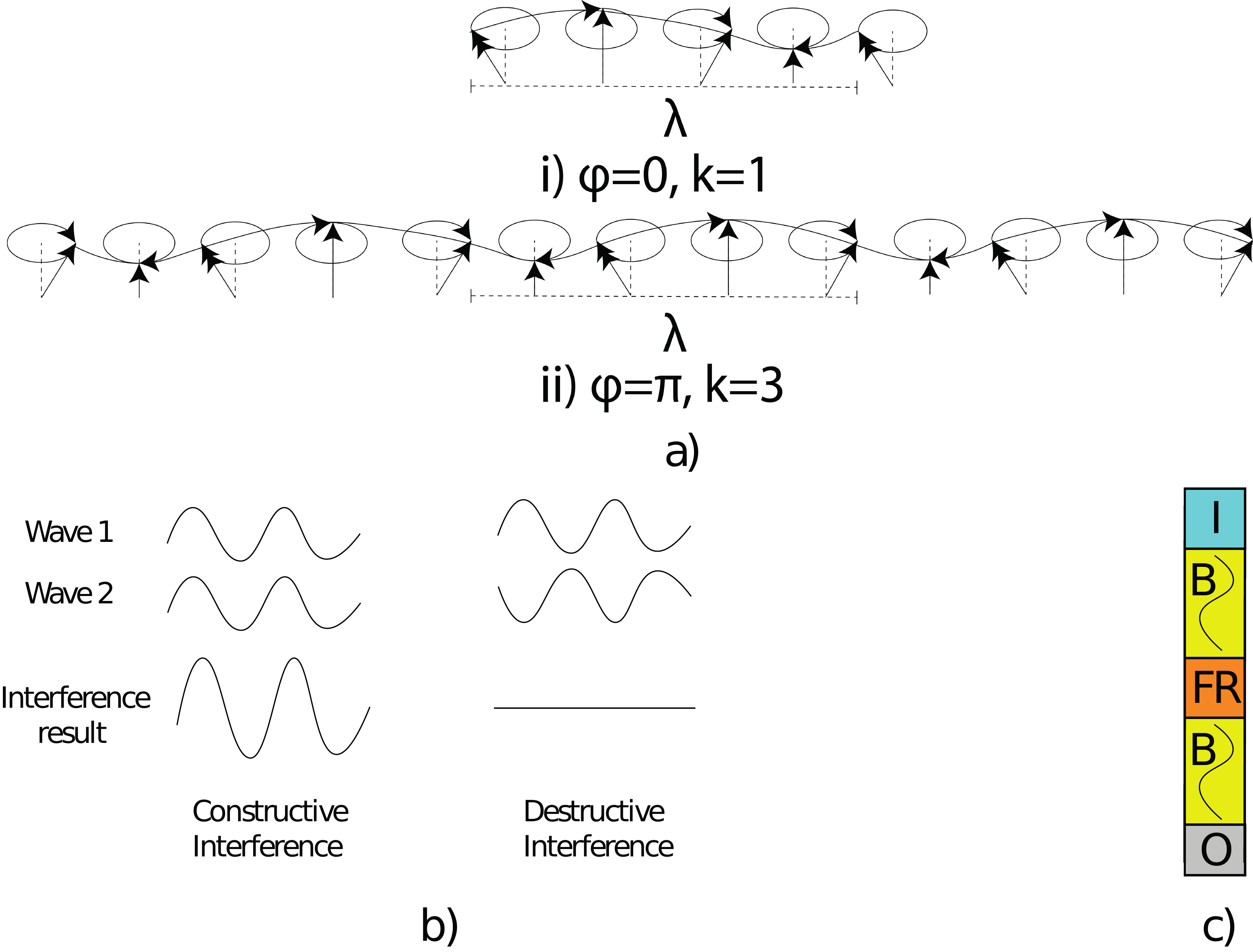}
  \caption{a) Constructive and Destructive Interference. b) Spin Wave Device.}
  \label{fig:spin_wave_device}
\end{figure}

Equation (\ref{eq:1}) has wave-like solutions under small magnetic disturbances, which are called Spin Waves (SWs) and are the collective excitations of the magnetization within the magnetic material \cite{amahmoud2}. A SW, as any other wave, is described by its amplitude $A$, phase $\phi$, wavelength $\lambda$, wavenumber $k=\frac{2\pi}{\lambda}$, and frequency $f$ as graphically presented in Figure \ref{fig:spin_wave_device}a) \cite{amahmoud2}. SW frequency and wavenumber are linked by the so called dispersion relation, which plays a fundamental role during the SW circuit design process \cite{amahmoud2}.

Generally speaking, information can be encoded in SW amplitude and phase at different frequencies \cite{amahmoud2,parallelism}, while the interaction between SWs coexisting in the same waveguide is governed by the interference principle. Figure \ref{fig:spin_wave_device}b) presents two SWs interaction situations: if they have the same phase, i.e., $\Delta \phi=0$, they interfere constructively resulting in a larger amplitude SW, whereas if they have different phases, i.e., $\Delta \phi=\pi$, they interfere destructively resulting in a diminished amplitude SW. Due to their very nature, SWs provide natural support for Majority function evaluation as the interference of an odd number of SWs emulates an Majority decision. For instance, if $3$ same amplitude, frequency, and wavelength SWs interfere, the result is a $0$ phase SW (logic $0$) if no more than one of them has a $\pi$ phase, and in a $\pi$ phase SW (logic $1$) otherwise, which is equivalent with the behavior of a 3-input Majority gate. Note that a CMOS 3-input majority gate implementation requires $18$ transistors, while in SW technology it only requires one waveguide. We note that if the SWs have different $A$, $\lambda$, and $f$, their interaction results in more sophisticated interferences, which might open different SW based computation paradigms. However, in this paper, we only consider the interaction of SWs with phase encoded information, i.e., logic $0$ and logic $1$ are represented by $0$ and $\pi$ phase, respectively. 

Figure \ref{fig:spin_wave_device}c) presents a SW device, which consists of an excitation region $I$, a waveguide $B$, a functional region $FR$, and a detection region $O$ \cite{amahmoud2,Magnonic_crystals_for_data_processing}. SW can be excited by means of, e.g., microstrip antennas, magnetoelectric cells, in region $I$ \cite{amahmoud2,Magnonic_crystals_for_data_processing}. $B$ can be made out of different materials such as Permalloy, Yttrium iron garnet, CoFeB, which must be properly chosen as it has a direct impact on the SW properties, and propagation \cite{amahmoud2,Magnonic_crystals_for_data_processing}. SWs can be amplified, normalized or interfere with other SWs within FR, and the output is detected at $O$ by using similar or different components than the one utilized in the excitation region \cite{amahmoud2,Magnonic_crystals_for_data_processing} by means of phase and threshold detection techniques \cite{amahmoud2}. Phase detection is based on comparing the resulted SW phase with a predefined phase. For instance, if the output SW has a phase difference $\Delta \phi=0$, the output is logic $0$, whereas the output is logic $1$ if the phase $\Delta \phi=\pi$. Threshold detection relies on the comparison of the output SW amplitude with a predefined threshold value $T$, i.e., if the SW amplitude is larger than $T$, the output is logic $1$, and $0$, otherwise\cite{amahmoud2,Magnonic_crystals_for_data_processing}.

\section{SW Approximate $4$:$2$ Compressor}
\label{sec:Proposed approximate functions}

\begin{figure}[t]
%\vspace{-0.5cm}
\centering
  \includegraphics[width=\linewidth]{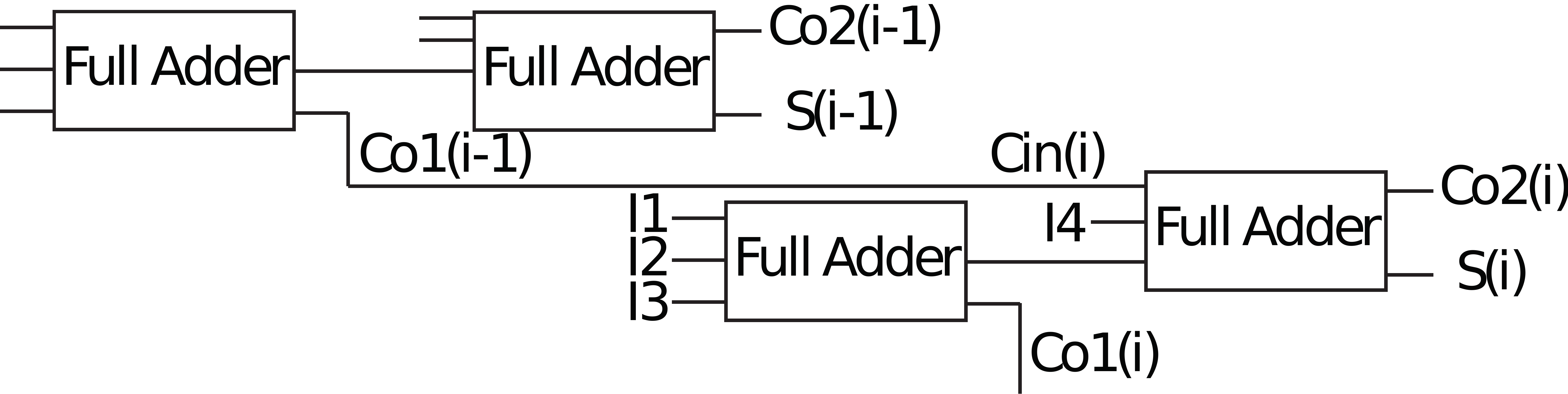}
  \caption{Conventional $4$:$2$ Compressor.}
  \label{fig:conventional_structure}
\end{figure}

For many state-of-the-art applications, e.g., artificial neural network, machine vision, detecting events such as visual surveillance and people counting, which heavily rely on multiplications the availability of fast multipliers is essential. Wallace or Dadda tree multipliers are the fastest and can perform a multiplication within $2$ clock cycles. They embed $3$ stages, i.e., partial product generation, reduction tree, and carry propagation adder. In an n-bit multiplier the first stage requires $n^2$ gates to produce the partial products matrix, the second stage provides a logarithmic depth reduction of $n$ $n$-bit numbers to two numbers without carry propagation, and the final stage is a carry propagate adder that sums-up the reduction tree outputs \cite{Prahami}. The $n$ to $2$ reduction has been traditionally done by means of Full and Half adders but $n$:$2$ compressors based reduction trees can be shallower and have a more regular layout \cite{Prahami}. Thus, most of the state-of-the-art CMOS implementations make use of $4$:$2$ compressors for which faster than $2$ cascaded FA implementations exists \cite{conventional_compressor1,conventional_compressor2,conventional_compressor4}. Essentially speaking, a $4$:$2$ compressor processes $4$ dots in the same column  and generate one dot in the current column and a carry to the next column. To properly preserve the value carried by the inputs, after a FA delay, the 4:2 compressor generates a transport to the next column and receives a transport from the previous position, which it further process
to generate the sum and a carry for the next column. Thus, the compressor has $5$ inputs (one of them coming from the previous column) and $2$ real outputs and one intermediate transport to the next column. Given that multiplication dominated error tolerant applications exist, e.g., multimedia processing and social media \cite{applications}, approximate CMOS $4$:$2$ compressors have been proposed \cite{conventional_compressor4}, which enable significant energy consumptions and area saving. 

Figure \ref{fig:conventional_structure} presents the conventional structure of an accurate SW $4$:$2$ compressor, which consists of $2$ full adders. When applied in column $i$ of the partial product matrix it processes $4$ dots in that column and a Carry-in $C_{in}$ reported by a $4$:$2$ compressor in column $i$-$1$, and generates $3$ outputs, $1$ intermediate transport $C_{o1}$ that serves as $C_{in}$ for a counter in column $i$+$1$, the Sum $S$ and Carry-out $C_{o2}$. A straightforward SW $4$:$2$ compressor implementation can be built using the SW full adder proposed in \cite{mahmoud2021spin}, which provides accurate results with acceptable delay and energy efficiency as further discussed in Section \ref{sec:Performance Evaluation}. However, as previously mentioned, many applications are error tolerant, and work properly within certain error limits \cite{applications}. Therefore, by enabling approximate computing, a more energy efficient SW $4$:$2$ compressor can be made.

\begin{figure}[t]
%\vspace{-0.5cm}
\centering
  \includegraphics[width=0.8\linewidth]{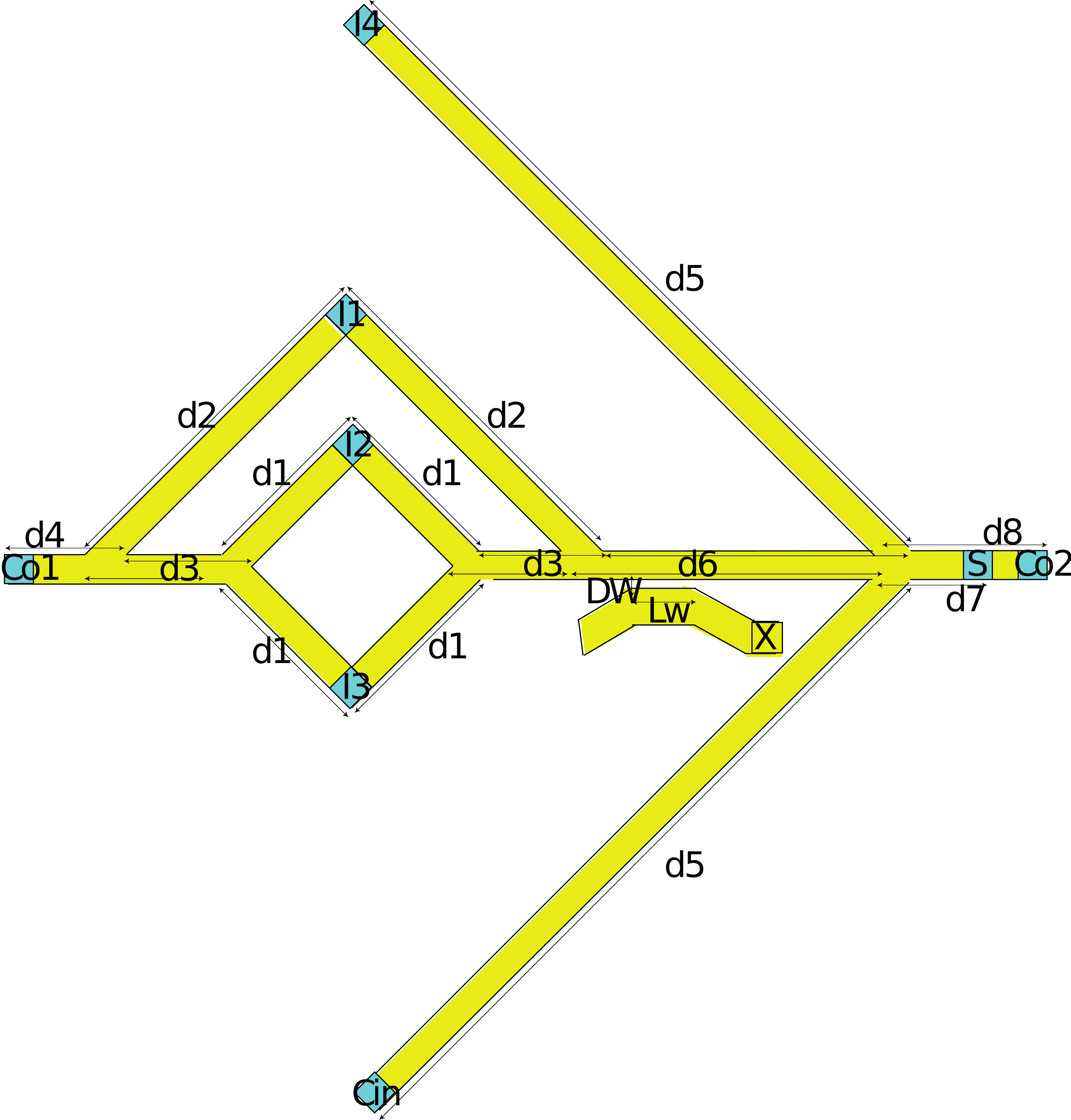}
  \caption{Approximate Spin Wave Based FA with Normalizer.}
  \label{fig:structure1}
\end{figure}

\begin{figure}[t]
%\vspace{-0.5cm}
\centering
  \includegraphics[width=0.8\linewidth]{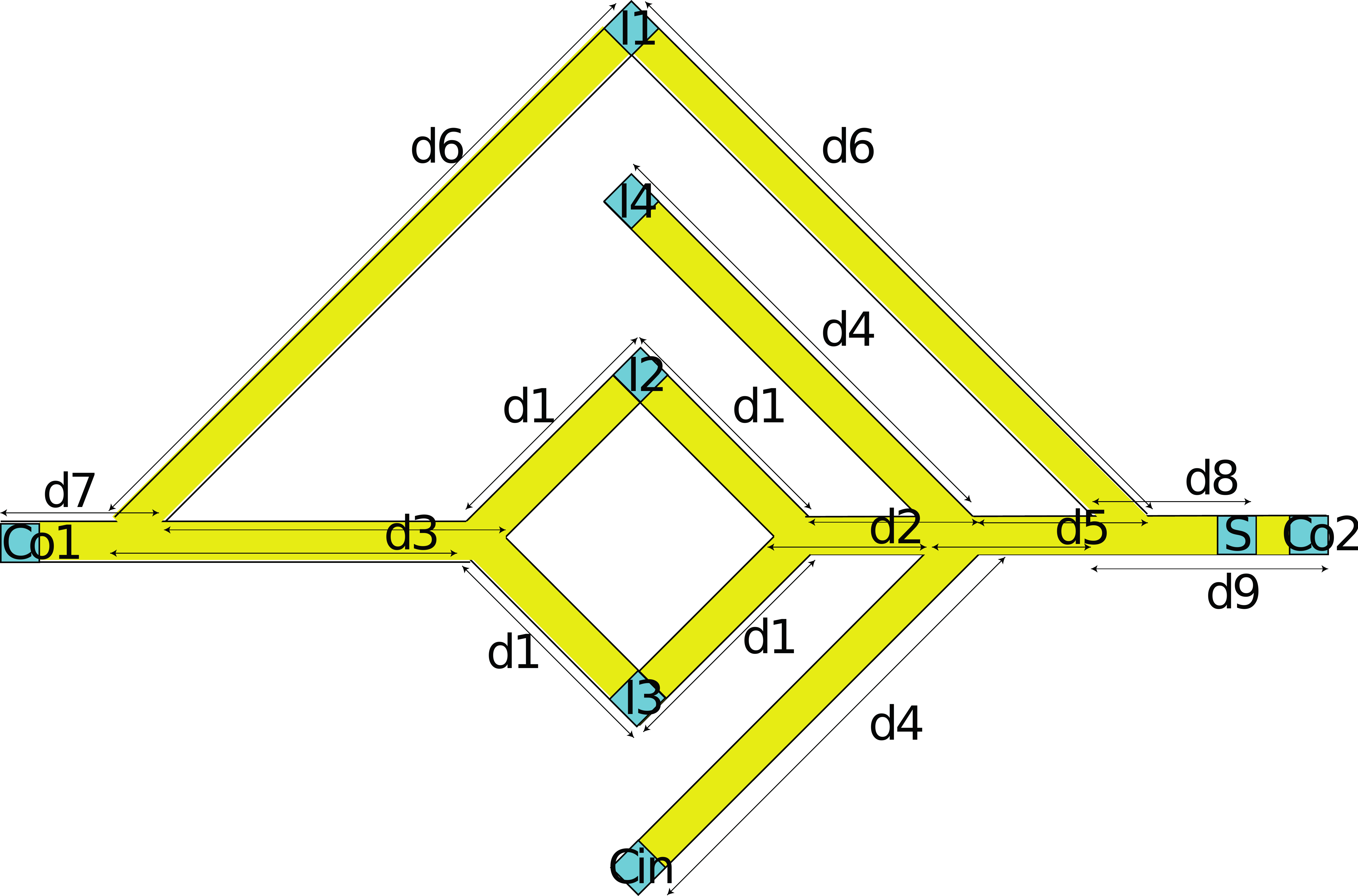}
  \caption{Approximate Spin Wave Based FA without Normalizer.}
  \label{fig:structure2}
\end{figure}

The straightforward implementation of a SW approximate $4$:$2$ compressor can be done by means of the two approximate SW full adder proposed in \cite{mahmoud2021spinapproximate}. This requires the cascading of two Full Adders (FAs), which cannot be performed straightforward because different FA input combinations generate different output SW strengths \cite{amahmoud1}. To solve this issue, and make the compressor functions correctly a directional coupler is required \cite{amahmoud1} to normalize the output of the first FA before passing it to the second FA. Figure \ref{fig:structure1} presents the approximate compressor obtained by cascading two approximate FAs by means of a normalizer (directional coupler). However, the directional coupler induces substantial delay and area overheads, which makes working without it desirable. Therefore, we propose the novel directional coupler free approximate compressor depicted in Figure \ref{fig:structure2}. The behaviour of the $2$ directly cascaded FAs is now obtained with a  $3$-input Majority gate and a $5$-input Majority gate computing $C_{o1}=MAJ(X,Y,C_i)$, and $S=\overline{C_{o2}}=\overline{MAJ(I_1,I_2,I_3,\overline{I_4},\overline{C_{in}})}$, respectively. 
The proposed $4$:$2$ approximate compressor generates $C_{o1}$ without any error, and $S$ and $C_{o2}$ with an average error rate of $31.25$\%, and $18.75$\%, respectively. Table \ref{table:1} presents the truth table of the accurate $4$:$2$ compressor $C_{o1}$, $S_{ac}$, and $C_{o2ac}$, the approximate $4$:$2$ compressor without directional coupler $C_{o1}$, $C_{o2ap1}$, and $S_{ap1}$, and the approximate $4$:$2$ compressor with directional coupler $C_{o1}$, $C_{o2ap2}$, and $S_{ap2}$. As it can be observed from the Table, approximate $4$:$2$ compressors with and without directional coupler provide the same average error rate of $25$\% because $S_{ap1}$, and $C_{o2ap1}$ have an error rate of $37.5$\%, and $12.5$\%, respectively, whereas $S_{ap1}$, and $C_{o2ap1}$ have an error rate of $31.25$\%, and $18.75$\%, respectively. Note that the erroneous outputs values in the Table are underlined and typeset in bold to highlight them.

\begin{table}[t]
\caption{Accurate and Approximate SW-based $4$:$2$ Compressor.}
\label{table:1}
\centering
\tiny
  \begin{tabular}{|c|c|c|c|c|c|c|c|}
    \hline
   $C_{in}I_4I_3I_2I_1$ & $C_{o1}$ & $C_{o2ac}$ & $C_{o2ap1}$ & $C_{o2ap2}$ & $S_{ac}$ & $S_{ap1}$ & $S_{ap2}$   \tabularnewline
    \hline
    $00000$ & $0$ & $0$ & $0$ & \textbf{\underline{1}} & $0$ & \textbf{\underline{1}} & $0$ \tabularnewline
    \hline
    $00001$ & $0$ & $0$ & $0$ & $0$ & $1$ & $1$ & $1$ \tabularnewline
    \hline
    $00010$ & $0$ & $0$ & $0$ & $0$ & $1$ & $1$ & $1$ \tabularnewline
    \hline
    $00011$ & $1$ & $0$ & $0$ & $0$ & $0$ & \textbf{\underline{1}} & \textbf{\underline{1}} \tabularnewline
    \hline
    $00100$ & $0$ & $0$ & $0$ & $0$ & $1$ & $1$ & $1$ \tabularnewline
    \hline
    $00101$ & $1$ & $0$ & $0$ & $0$ & $0$ & \textbf{\underline{1}} & \textbf{\underline{1}} \tabularnewline
    \hline
    $00110$ & $1$ & $0$ & $0$ & $0$ & $0$ & \textbf{\underline{1}} & \textbf{\underline{1}} \tabularnewline
    \hline
    $00111$ & $1$ & $0$ & $0$ & $0$ & $1$ & $1$ & $1$  \tabularnewline
    \hline
    $01000$ & $0$ & $0$ & \textbf{\underline{1}} & \textbf{\underline{1}} & $1$ & \textbf{\underline{0}} & \textbf{\underline{0}} \tabularnewline
    \hline
    $01001$ & $0$ & $1$ & $1$ & $1$ & $0$ & $0$ & $0$ \tabularnewline
    \hline
    $01010$ & $0$ & $1$ & $1$ & $1$ & $0$ & $0$ & $0$ \tabularnewline
    \hline
    $01011$ & $1$ & $0$ & $0$ & $0$ & $1$ & $1$ & $1$  \tabularnewline
    \hline
    $01100$ & $0$ & $1$ & $1$ & $1$ & $0$ & $0$ & $0$ \tabularnewline
    \hline
    $01101$ & $1$ & $0$ & $0$ & $0$ & $1$ & $1$ & $1$ \tabularnewline
    \hline
    $01110$ & $1$ & $0$ & $0$ & $0$ & $1$ & $1$ & $1$ \tabularnewline
    \hline
    $01111$ & $1$ & $1$ & \textbf{\underline{0}} & \textbf{\underline{0}} & $0$ & \textbf{\underline{1}} & \textbf{\underline{1}} \tabularnewline
    \hline
    $10000$ & $0$ & $0$ & \textbf{\underline{1}} & \textbf{\underline{1}} & $1$ & \textbf{\underline{0}} & \textbf{\underline{0}} \tabularnewline
    \hline
    $10001$ & $0$ & $1$ & $1$ & $1$ & $0$ & $0$ & $0$ \tabularnewline
    \hline
    $10010$ & $0$ & $1$ & $1$ & $1$ & $0$ & $0$ & $0$ \tabularnewline
    \hline
    $10011$ & $1$ & $0$ & $0$ & $0$ & $1$ & $1$ & $1$ \tabularnewline
    \hline
    $10100$ & $0$ & $1$ & $1$ & $1$ & $0$ & $0$ & $0$ \tabularnewline
    \hline
    $10101$ & $1$ & $0$ & $0$ & $0$ & $1$ & $1$ & $1$ \tabularnewline
    \hline
    $10110$ & $1$ & $0$ & $0$ & $0$ & $1$ & $1$ & $1$ \tabularnewline
    \hline
    $10111$ & $1$ & $1$ & \textbf{\underline{0}} & \textbf{\underline{0}} & $0$ & \textbf{\underline{1}} & \textbf{\underline{1}} \tabularnewline
    \hline
    $11000$ & $0$ & $1$ & $1$ & $1$ & $0$ & $0$ & $0$ \tabularnewline
    \hline
    $11001$ & $0$ & $1$ & $1$ & $1$ & $1$ & \textbf{\underline{0}} & \textbf{\underline{0}} \tabularnewline
    \hline
    $11010$ & $0$ & $1$ & $1$ & $1$ & $1$ & \textbf{\underline{0}} & \textbf{\underline{0}} \tabularnewline
    \hline
    $11011$ & $1$ & $1$ & $1$ & $1$ & $0$ & $0$ & $0$ \tabularnewline
    \hline
    $11100$ & $0$ & $1$ & $1$ & $1$ & $1$ & \textbf{\underline{0}} & \textbf{\underline{0}} \tabularnewline
    \hline
    $11101$ & $1$ & $1$ & $1$ & $1$ & $0$ & $0$ & $0$ \tabularnewline
    \hline
    $11110$ & $1$ & $1$ & $1$ & $1$ & $0$ & $0$ & $0$ \tabularnewline
    \hline
    $11111$ & $1$ & $1$ & $1$ & \textbf{\underline{0}} & $1$ & \textbf{\underline{0}} & $1$ \tabularnewline
    \hline
    \end{tabular}
\end{table}

To achieve proper functionality for the structure in Figure \ref{fig:structure2}, the waveguide width must be smaller or equal to the SW wavelength to simplify the interference patterns, all SWs must be excited at the same amplitude, wavelength, and frequency, and the waveguide lengths must be accurately computed as they determine the SWs interaction modes. For example, if SW constructive (destructive) interference is envisaged for in phase (out of phase) SWs, the distances must be equal with $n \times \lambda$, where $n=0,1,2,\ldots$; this is the case for $d_1$, $d_3$, $d_4$, and $d_6$ in Figure \ref{fig:structure2}. In contrast, if SW constructive (destructive) interference is envisaged for out of phase (in phase) SWs, the distances must be equal with $(n+1/2) \times \lambda$; this is the case for $d_2$ and $d_5$ in Figure \ref{fig:structure2}. On the output side, it is important to detect the output at specific position, i.e., if the desired output is the output itself, which is the case for $C_{o1}$ in Figure \ref{fig:structure2}, $d_7$ must be equal with $n \times \lambda$, whereas if the inverted output is desired, the distance must be equal with $(n+1)/2 \times \lambda$. Moreover, the outputs must be detected as near as possible from the last interference point to capture large SW amplitude.

The proposed SW $4$:$2$ compressor operation principle is as follows: 
\begin{itemize}
  \item $C_{o1}$: SWs are excited at $I_1$, $I_2$, and $I_3$ with the same amplitude, wavelength, and frequency at the same time moment. The $I_2$ SW interfere constructively or destructively with $I_3$ SW depending on their phase difference, the resulted SW propagates through the waveguide, and subsequently interferes with the $I_1$ SW. The resulted SW is captured at the output $C_{o1}$ based on phase detection. 
  \item $S$ and $C_{o2}$: $I_2$ SW interferes constructively or destructively with $I_3$ SW depending on their phase difference, and the resulted SW propagates through the waveguide to interfere with the SWs excited at $I_4$ and $C_{in}$. The resulted SW propagates, and subsequently interferes with the $I_1$ SW. Finally, the resulted SW is captured at the outputs $S$ and $C_{o2}$ based on the threshold detection. 
\end{itemize}

\section{Simulation Setup and Results}
\label{sec:Simulation Setup and Results}

In order to validate the proposed structure by MuMax3 \cite{mumax}, we made use of the parameters specified in Table \ref{table:3} \cite{parameters}. In addition, we assumed waveguide thickness and width of \SI{1}{nm} and \SI{50}{nm}, respectively, to guarantee high SW group velocity. Furthermore, we excite the SWs with Gaussian pulses with \SI{500}{ps} sigma modulated at \SI{10}{GHz} to save energy, gaurantee the excitation of single frequency SWs, and achieve high group velocity. From the SW dispersion relation, at \SI{10}{GHz}, we determine $k$ as being \SI{36.9}{rad/\mu m}, which results in a $\lambda$ = $2 \pi/k$ = \SI{170}{nm}. As discussed in Section \ref{sec:Proposed approximate functions}, the distances $d_1$, $d_2$, \ldots, $d_8$ should be equal to integer multiples of $\lambda$, and are: $d_1$ = \SI{170}{nm} (n = 1), $d_2$ = \SI{595}{nm} (n = 3.5), $d_3$ = \SI{1190}{nm} (n = 7), $d_4$ = \SI{510}{nm} (n = 2), $d_5$ = \SI{595}{nm} (n = 3.5), $d_6$ = \SI{1190}{nm} (n = 7), $d_7$ = \SI{170}{nm} (n = 1), $d_8$ = \SI{85}{nm} (n = 0), and $d_9$ = \SI{170}{nm} (n = 1).

\begin{table}[t]
\caption{Simulation Parameters.}
\label{table:2}
\centering
  \begin{tabular}{|c|c|}
    \hline
    Parameters & Values \\
    \hline
    Waveguide Material & $Fe_{60}Co_{20}B_{20}$ \\
    \hline
    Saturation magnetization $M_s$ & $1.1$ $\times$ $10^6$ A/m \\
    \hline
    Perpendicular anisotropy constant $k_{ani}$ & $0.83$ MJ/$m^3$\\
    \hline
    Damping constant $\alpha$ & $0.004$ \\
    \hline
    Exchange stiffness $A_{exch}$ & $18.5$ pJ/m \\
    \hline
  \end{tabular}
\end{table}

\begin{figure}[t]
\centering
  \includegraphics[width=\linewidth]{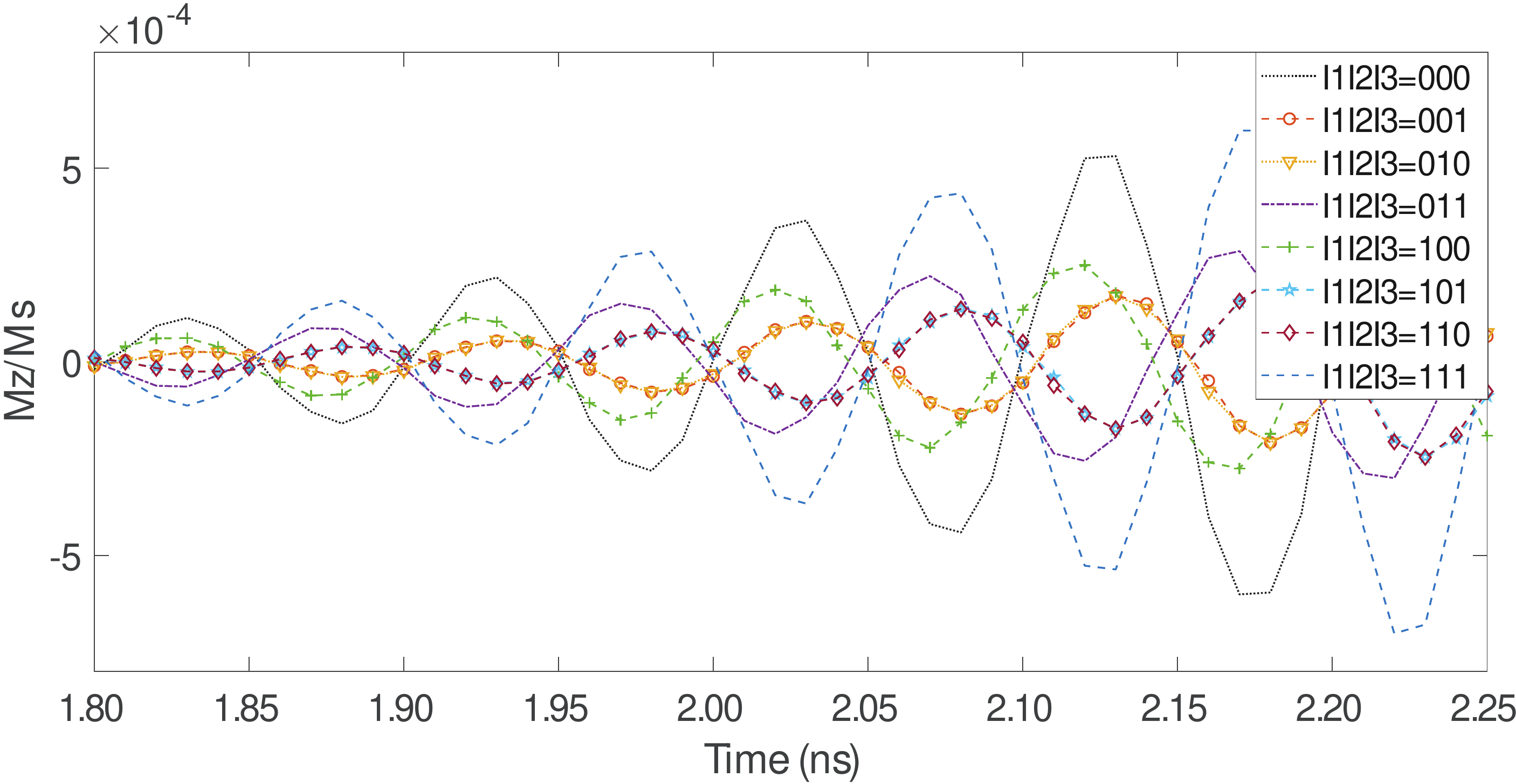}
  \caption{Normalized $4$:$2$ Compressor Output $C_{o1}$.}
  \label{fig:result}
\end{figure}

Figure \ref{fig:result} presents $C_{o1}$ MuMax3 simulation results for \{$I_1$,$I_2$,$I_3$\} = \{$0$,$0$,$0$\}, \{$0$,$0$,$0$\}, \{$0$,$0$,$1$\}, \{$0$,$1$,$0$\}, \{$0$,$1$,$1$\}, \{$1$,$0$,$0$\}, \{$1$,$0$,$1$\}, \{$1$,$1$,$0$\}, and \{$1$,$1$,$1$\}. One can observe in the Figure that $C_{o1}$ is detected correctly. $C_{o1}=0$ for  \{$I_1$,$I_2$,$I_3$\}= \{$0$,$0$,$0$\}, \{$0$,$0$,$1$\}, \{$0$,$1$,$0$\}, and \{$0$,$1$,$1$\}, whereas $C_{o1}=1$ for  \{$I_1$,$I_2$,$I_3$\} = \{$1$,$0$,$0$\}, \{$1$,$0$,$1$\}, \{$1$,$1$,$0$\}, and \{$1$,$1$,$1$\}, as it should, for a \SI{0.4}{ns} reading window  starting \SI{1.80}{ns} after the input application. 

%identify it on figure 1

Table \ref{table:3} presents the normalized magnetization of the $4$:$2$ approximate compressor outputs $C_{o2}$ and $S$ for all possible input combinations, i.e., \{$C_{in}$,$I_4$,$I_3$,$I_2$,$I_1$\}= \{$0$,$0$,$0$,$0$,$0$\}, \{$0$,$0$,$0$,$0$,$1$\}, \ldots, \{$1$,$1$,$1$,$1$,$0$\}, and \{$1$,$1$,$1$,$1$,$1$\}. Note that threshold technique is used to detect $C_{o2}$, and $S$, i.e., if the normalized magnetization of the output SW is larger than the threshold, value $T$, the output is logic $1$ and $0$, otherwise. For $C_{o2}$ detection $T=0$ is appropriate, which results in $C_{o2}=1$ for input combinations \{$C_{in}$,$I_4$,$I_3$,$I_2$,$I_1$\}= \{$0$,$0$,$0$,$0$,$0$\}, \{$0$,$1$,$0$,$0$,$0$\}, \{$0$,$1$,$0$,$0$,$1$\}, \{$0$,$1$,$0$,$1$,$0$\}, \{$0$,$1$,$1$,$0$,$0$\}, \{$1$,$0$,$0$,$0$,$0$\}, \{$1$,$0$,$0$,$0$,$1$\}, \{$1$,$0$,$0$,$1$,$0$\}, \{$1$,$0$,$1$,$0$,$0$\}, \{$1$,$1$,$0$,$0$,$0$\}, \{$1$,$1$,$0$,$0$,$1$\}, \{$1$,$1$,$0$,$1$,$0$\}, \{$1$,$1$,$0$,$1$,$1$\}, \{$1$,$1$,$1$,$0$,$0$\}, \{$1$,$1$,$1$,$0$,$1$\}, and \{$1$,$1$,$1$,$1$,$0$\}, and $C_{o2}=0$ for the remaining cases, as it should.

The same threshold value is suitable for $S$, but the threshold condition is flipped, i.e., if the resulted SW normalized magnetization is larger than $0$, $S$ is logic $0$, and logic $1$, otherwise. This results in $S=0$ for \{$C_{in}$,$I_4$,$I_3$,$I_2$,$I_1$\}= \{$0$,$0$,$0$,$0$,$0$\}, \{$0$,$1$,$0$,$0$,$0$\}, \{$0$,$1$,$0$,$0$,$1$\}, \{$0$,$1$,$0$,$1$,$0$\}, \{$0$,$1$,$1$,$0$,$0$\}, \{$1$,$0$,$0$,$0$,$0$\}, \{$1$,$0$,$0$,$0$,$1$\}, \{$1$,$0$,$0$,$1$,$0$\}, \{$1$,$0$,$1$,$0$,$0$\}, \{$1$,$1$,$0$,$0$,$0$\}, \{$1$,$1$,$0$,$0$,$1$\}, \{$1$,$1$,$0$,$1$,$0$\}, \{$1$,$1$,$0$,$1$,$1$\}, \{$1$,$1$,$1$,$0$,$0$\}, \{$1$,$1$,$1$,$0$,$1$\}, and \{$1$,$1$,$1$,$1$,$0$\}, and $S=1$ for the remaining cases, as it should. 

Therefore, the MuMax3 simulations proves that the proposed $4$:$2$ approximate compressor provides the expected functionality.

\begin{table}[t]
\caption{Normalized Approximate SW-based $4$:$2$ Compressor Outputs $C_{o2}$ and $S$.}
\label{table:3}
\centering
\scalebox{0.85}{%
\begin{tabular}{|c|c|c|c|}
    \hline
   $C_{in}I_4I_3I_2I_1$ & Resulting SW  & $C_{o2}$ after thresholding & $S$ after thresholding   \tabularnewline
    \hline
    $00000$ & $0.45$ & $1$ & $0$ \tabularnewline
    \hline
    $00001$ & $-0.08$ & $0$ & $1$ \tabularnewline
    \hline
    $00010$ & $-0.07$ & $0$ & $1$ \tabularnewline
    \hline
    $00011$ & $-0.59$ & $0$ & $1$ \tabularnewline
    \hline
    $00100$ & $-0.01$ & $0$ & $1$ \tabularnewline
    \hline
    $00101$ & $-0.46$ & $0$ & $1$ \tabularnewline
    \hline
    $00110$ & $-0.49$ & $0$ & $1$ \tabularnewline
    \hline
    $00111$ & $-1$ & $0$ & $1$  \tabularnewline
    \hline
    $01000$ & $0.66$ & $1$ & $0$ \tabularnewline
    \hline
    $01001$ & $0.23$ & $1$ & $0$ \tabularnewline
    \hline
    $01010$ & $0.22$ & $1$ & $0$ \tabularnewline
    \hline
    $01011$ & $-0.3$ & $0$ & $1$  \tabularnewline
    \hline
    $01100$ & $0.3$ & $1$ & $0$ \tabularnewline
    \hline
    $01101$ & $-0.21$ & $0$ & $1$ \tabularnewline
    \hline
    $01110$ & $-0.2$ & $0$ & $1$ \tabularnewline
    \hline
    $01111$ & $-0.69$ & $0$ & $1$ \tabularnewline
    \hline
    $10000$ & $0.68$ & $1$ & $0$ \tabularnewline
    \hline
    $10001$ & $0.18$ & $1$ & $0$ \tabularnewline
    \hline
    $10010$ & $0.21$ & $1$ & $0$ \tabularnewline
    \hline
    $10011$ & $-0.28$ & $0$ & $1$ \tabularnewline
    \hline
    $10100$ & $0.28$ & $1$ & $0$ \tabularnewline
    \hline
    $10101$ & $-0.22$ & $0$ & $1$ \tabularnewline
    \hline
    $10110$ & $-0.18$ & $0$ & $1$ \tabularnewline
    \hline
    $10111$ & $-0.73$ & $0$ & $1$ \tabularnewline
    \hline
    $11000$ & $1$ & $1$ & $0$ \tabularnewline
    \hline
    $11001$ & $0.51$ & $1$ & $0$ \tabularnewline
    \hline
    $11010$ & $0.47$ & $1$ & $0$ \tabularnewline
    \hline
    $11011$ & $0.012$ & $1$ & $0$ \tabularnewline
    \hline
    $11100$ & $0.59$ & $1$ & $0$ \tabularnewline
    \hline
    $11101$ & $0.07$ & $1$ & $0$ \tabularnewline
    \hline
    $11110$ & $0.09$ & $1$ & $0$ \tabularnewline
    \hline
    $11111$ & $-0.4$ & $0$ & $1$  \tabularnewline
    \hline
    \end{tabular} }
\end{table}

\section{Performance Evaluation and Discussion}
\label{sec:Performance Evaluation}

We evaluate the proposed SW approximate $4$:$2$ compressor and compare it in terms of error rate, energy consumption, delay, and area (the number of utilized devices) with the state-of-the-art SW, \SI{45}{nm} CMOS \cite{approxCompCMOS1}, and Spin-CMOS \cite{SPIN} counterparts. In order to assess the performance of our proposal, we make the following assumptions: (i) Magnetoelectric (ME) cells having a power consumption of \SI{34}{nW}, and a delay of \SI{0.42}{ns} \cite{Excitation_table_ref16} are utilized for SW excitation/detection. (ii) SWs consume negligible energy during interference and propagation through waveguides. Note that these assumptions might need to be revisited to better capture SW technology future developments.   

The proposed compressor with and without Directional Coupler (DC) delays can be calculated by adding the SW propagation determined by means of micro-magnetic simulations, and the delay of the excitation and detection cells, which sums-up to \SI{11.4}{ns} and \SI{3.4}{ns}, respectively. We note that in order to perform amplitude normalization the DC has to be rather long \cite{amahmoud1}, which results in a large delay overhead. 

Table \ref{table:4} presents the evaluation results. When compared with the accurate SW compressor, which is a direct implementation consisting of two accurate SW adders in \cite{mahmoud2021spin}, the proposed $4$:$2$ compressor saves $31.5$\% energy and is $1.93$x faster. Moreover, it has the same energy consumption, and error rate as the approximate compressor with DC, but it requires $3$x less delay. In addition, it consumes $20$\%  and $14$\% less energy, has approximately $2$ orders of magnitude higher delay, and exhibits $61$\% more and $17$\% less average error rate when compared with CMOS1 and CMOS2 designs in Table \ref{table:4}, respectively. When compared with same error rate Spin-CMOS (Spin-CMOS1 design in Table \ref{table:4}), it consumes $3$ orders of magnitude less energy and provides a $17$\% delay reduction. Although Spin-CMOS2 design provides $19$\% better average error rate, it is $3$ order of magnitude less effective in terms of energy consumption and slower. Note that the proposed compressor requires the smallest number of devices, which indicates that it potentially requires the lowest chip real-estate.  .

\begin{table}[t]
\caption{Approximate $4$:$2$ Compressor Performance Comparison.}
\label{table:4}
\centering
  \begin{tabular}{|c|c|c|c|c|c|}
    \hline
    Technology & Type & Error Rate & Energy (fJ)  &  Delay (ns) & Device No. \tabularnewline
    \hline
    Spin Wave & Accurate  & $0$ & $0.2$  & $6.56$ & $14$ \tabularnewline \hline
    Spin Wave  (with DC) & Approximate  & $0.31$ & $0.137$  & $11.4$ & $8$ \tabularnewline \hline
    Spin Wave (without DC) & Approximate  & $0.31$ & $0.137$  & $3.4$ & $8$ \tabularnewline \hline
    CMOS1 \cite{approxCompCMOS1}& Approximate  & $0.125$  & $0.172$ & $0.049$ & $40$ \tabularnewline \hline
    CMOS2 \cite{approxCompCMOS1}& Approximate  & $0.375$ & $0.16$  & $0.048$ & $28$ \tabularnewline \hline
    Spin-CMOS1 \cite{SPIN}& Approximate  & $0.31$ & $173$  & $3$ & $28$ \tabularnewline \hline
    Spin-CMOS2 \cite{SPIN}& Approximate  & $0.25$ & $338$  & $4$ & $42$ \tabularnewline \hline
  \end{tabular}
\end{table}

To get some inside on the implications of our proposal at the application level, we consider the well-known JPEG encoding, which makes use the Discrete Cosine Transform (DCT) , \cite{JPEG} as discussion vehicle. Given that JPEG encoding is error tolerant and DCT is a multiplication dominated algorithm, 4:2 approximate compressors based tree multipliers are quite attractive for practical JPEG codec implementations. Such an approach has been presented in \cite{SPIN} and given that the approximate $4$:$2$ compressor in \cite{SPIN} has the same average error rate as the one we propose, we can infer that replacing their compressor with ours does not change the image quality while resulting with $3$ orders of magnitude less energy consumption.

We note that the main goal of this paper is to propose and validate a SW $4$:$2$ approximate compressor and as such we do not take into consideration thermal and variability effects. However, in \cite{DC}, it was suggested that thermal noise, edge roughness, and waveguide trapezoidal cross section do not have noticeable impact on gate functionality. Thus, we expect that the $4$:$2$ approximate compressor functions correctly under their presence. However, further investigation of such phenomena is of great interest but cannot be performed before technology data and suitable simulation tools become available.

\section{Conclusions}
\label{sec:Conclusion}
This paper proposed a Spin Wave (SW) based $4$:$2$ approximate compressor, which consists of $3$-input and $5$-input Majority gates. We reported the design of approximate circuits without directional couplers, which are essential to normalize gate output(s) when cascading them in accurate circuit designs. We validated the proposed compressor by means of micromagnetic simulations, and compared it with the state-of-the-art SW, \SI{22}{nm} CMOS, \SI{45}{nm} CMOS, and Spin-CMOS counterparts. 

The evaluation results indicated that the proposed $4$:$2$ compressor saves $31.5$\% energy in comparison with the accurate SW compressor, has the same energy consumption, and error rate as the approximate compressor with DC, but it required $3$x less delay. Moreover, it consumes $14$\% less energy, while having $17$\% lower error rate when compared with the approximate \SI{45}{nm} CMOS counterpart. Furthermore, it outperformes the approximate Spin-CMOS based compressor by $3$ orders of magnitude in term of energy consumption while providing the same error rate. Last but not least, the proposed compressor requires the smallest number of devices, thus it potentially requires the lowest chip real-estate.

\section*{Acknowledgement}
This work has received funding from the European Union's Horizon 2020 research and innovation program within the FET-OPEN project CHIRON under grant agreement No. 801055. It has also been partially supported by imec's industrial affiliate program on beyond-CMOS logic. F.V. acknowledges financial support from Flanders Research Foundation (FWO) through grant No.~1S05719N.

\bibliography{Spin_Wave_Based_Approximate_4_2_Compressor}

\begin{thebibliography}{31}%
\makeatletter
\providecommand \@ifxundefined [1]{%
 \@ifx{#1\undefined}
}%
\providecommand \@ifnum [1]{%
 \ifnum #1\expandafter \@firstoftwo
 \else \expandafter \@secondoftwo
 \fi
}%
\providecommand \@ifx [1]{%
 \ifx #1\expandafter \@firstoftwo
 \else \expandafter \@secondoftwo
 \fi
}%
\providecommand \natexlab [1]{#1}%
\providecommand \enquote  [1]{``#1''}%
\providecommand \bibnamefont  [1]{#1}%
\providecommand \bibfnamefont [1]{#1}%
\providecommand \citenamefont [1]{#1}%
\providecommand \href@noop [0]{\@secondoftwo}%
\providecommand \href [0]{\begingroup \@sanitize@url \@href}%
\providecommand \@href[1]{\@@startlink{#1}\@@href}%
\providecommand \@@href[1]{\endgroup#1\@@endlink}%
\providecommand \@sanitize@url [0]{\catcode `\\12\catcode `\$12\catcode
  `\&12\catcode `\#12\catcode `\^12\catcode `\_12\catcode `\%12\relax}%
\providecommand \@@startlink[1]{}%
\providecommand \@@endlink[0]{}%
\providecommand \url  [0]{\begingroup\@sanitize@url \@url }%
\providecommand \@url [1]{\endgroup\@href {#1}{\urlprefix }}%
\providecommand \urlprefix  [0]{URL }%
\providecommand \Eprint [0]{\href }%
\providecommand \doibase [0]{http://dx.doi.org/}%
\providecommand \selectlanguage [0]{\@gobble}%
\providecommand \bibinfo  [0]{\@secondoftwo}%
\providecommand \bibfield  [0]{\@secondoftwo}%
\providecommand \translation [1]{[#1]}%
\providecommand \BibitemOpen [0]{}%
\providecommand \bibitemStop [0]{}%
\providecommand \bibitemNoStop [0]{.\EOS\space}%
\providecommand \EOS [0]{\spacefactor3000\relax}%
\providecommand \BibitemShut  [1]{\csname bibitem#1\endcsname}%
\let\auto@bib@innerbib\@empty
%</preamble>
\bibitem [{\citenamefont {Shah}, \citenamefont {Steyerberg},\ and\
  \citenamefont {Kent}(2018)}]{data1}%
  \BibitemOpen
  \bibfield  {author} {\bibinfo {author} {\bibfnamefont {N.~D.}\ \bibnamefont
  {Shah}}, \bibinfo {author} {\bibfnamefont {E.~W.}\ \bibnamefont
  {Steyerberg}}, \ and\ \bibinfo {author} {\bibfnamefont {D.~M.}\ \bibnamefont
  {Kent}},\ }\href {\doibase 10.1001/jama.2018.5602} {\bibfield  {journal}
  {\bibinfo  {journal} {JAMA}\ }\textbf {\bibinfo {volume} {320}},\ \bibinfo
  {pages} {27} (\bibinfo {year} {2018})},\ \Eprint
  {http://arxiv.org/abs/https://jamanetwork.com/journals/jama/articlepdf/2683125/jama\_shah\_2018\_vp\_180051.pdf}
  {https://jamanetwork.com/journals/jama/articlepdf/2683125/jama\_shah\_2018\_vp\_180051.pdf}
  \BibitemShut {NoStop}%
\bibitem [{\citenamefont {Haron}\ and\ \citenamefont
  {Hamdioui}(2008)}]{cmosscaling1}%
  \BibitemOpen
  \bibfield  {author} {\bibinfo {author} {\bibfnamefont {N.~Z.}\ \bibnamefont
  {Haron}}\ and\ \bibinfo {author} {\bibfnamefont {S.}~\bibnamefont
  {Hamdioui}},\ }in\ \href {\doibase 10.1109/IDT.2008.4802475} {\emph {\bibinfo
  {booktitle} {2008 3rd International Design and Test Workshop}}}\ (\bibinfo
  {year} {2008})\ pp.\ \bibinfo {pages} {98--103}\BibitemShut {NoStop}%
\bibitem [{\citenamefont {Jiang}\ \emph {et~al.}(2019)\citenamefont {Jiang},
  \citenamefont {Laurenciu}, \citenamefont {Wang},\ and\ \citenamefont
  {Cotofana}}]{Yande1}%
  \BibitemOpen
  \bibfield  {author} {\bibinfo {author} {\bibfnamefont {Y.}~\bibnamefont
  {Jiang}}, \bibinfo {author} {\bibfnamefont {N.~C.}\ \bibnamefont
  {Laurenciu}}, \bibinfo {author} {\bibfnamefont {H.}~\bibnamefont {Wang}}, \
  and\ \bibinfo {author} {\bibfnamefont {S.~D.}\ \bibnamefont {Cotofana}},\
  }\href {\doibase 10.1109/TNANO.2019.2903480} {\bibfield  {journal} {\bibinfo
  {journal} {IEEE Transactions on Nanotechnology}\ }\textbf {\bibinfo {volume}
  {18}},\ \bibinfo {pages} {287} (\bibinfo {year} {2019})}\BibitemShut
  {NoStop}%
\bibitem [{\citenamefont {Nguyen}\ \emph {et~al.}(2020)\citenamefont {Nguyen},
  \citenamefont {Yu}, \citenamefont {Lebdeh}, \citenamefont {Taouil},
  \citenamefont {Hamdioui},\ and\ \citenamefont {Catthoor}}]{memristor10}%
  \BibitemOpen
  \bibfield  {author} {\bibinfo {author} {\bibfnamefont {H.~A.~D.}\
  \bibnamefont {Nguyen}}, \bibinfo {author} {\bibfnamefont {J.}~\bibnamefont
  {Yu}}, \bibinfo {author} {\bibfnamefont {M.~A.}\ \bibnamefont {Lebdeh}},
  \bibinfo {author} {\bibfnamefont {M.}~\bibnamefont {Taouil}}, \bibinfo
  {author} {\bibfnamefont {S.}~\bibnamefont {Hamdioui}}, \ and\ \bibinfo
  {author} {\bibfnamefont {F.}~\bibnamefont {Catthoor}},\ }\href {\doibase
  10.1145/3365837} {\bibfield  {journal} {\bibinfo  {journal} {J. Emerg.
  Technol. Comput. Syst.}\ }\textbf {\bibinfo {volume} {16}} (\bibinfo {year}
  {2020}),\ 10.1145/3365837}\BibitemShut {NoStop}%
\bibitem [{\citenamefont {Agarwal}\ \emph {et~al.}(2018)\citenamefont {Agarwal}
  \emph {et~al.}}]{ITRS}%
  \BibitemOpen
  \bibfield  {author} {\bibinfo {author} {\bibfnamefont {S.}~\bibnamefont
  {Agarwal}} \emph {et~al.},\ }\href@noop {} {\enquote {\bibinfo {title}
  {International roadmap of devices and systems 2017 edition: Beyond cmos
  chapter.}}\ }\bibinfo {type} {Tech. Rep.}\ (\bibinfo  {institution} {Sandia
  National Lab.(SNL-NM), United States},\ \bibinfo {year} {2018})\BibitemShut
  {NoStop}%
\bibitem [{\citenamefont {Mahmoud}\ \emph
  {et~al.}(2020{\natexlab{a}})\citenamefont {Mahmoud}, \citenamefont
  {Ciubotaru}, \citenamefont {Vanderveken}, \citenamefont {Chumak},
  \citenamefont {Hamdioui}, \citenamefont {Adelmann},\ and\ \citenamefont
  {Cotofana}}]{amahmoud2}%
  \BibitemOpen
  \bibfield  {author} {\bibinfo {author} {\bibfnamefont {A.}~\bibnamefont
  {Mahmoud}}, \bibinfo {author} {\bibfnamefont {F.}~\bibnamefont {Ciubotaru}},
  \bibinfo {author} {\bibfnamefont {F.}~\bibnamefont {Vanderveken}}, \bibinfo
  {author} {\bibfnamefont {A.~V.}\ \bibnamefont {Chumak}}, \bibinfo {author}
  {\bibfnamefont {S.}~\bibnamefont {Hamdioui}}, \bibinfo {author}
  {\bibfnamefont {C.}~\bibnamefont {Adelmann}}, \ and\ \bibinfo {author}
  {\bibfnamefont {S.}~\bibnamefont {Cotofana}},\ }\href {\doibase
  10.1063/5.0019328} {\bibfield  {journal} {\bibinfo  {journal} {Journal of
  Applied Physics}\ }\textbf {\bibinfo {volume} {128}},\ \bibinfo {pages}
  {161101} (\bibinfo {year} {2020}{\natexlab{a}})},\ \Eprint
  {http://arxiv.org/abs/https://doi.org/10.1063/5.0019328}
  {https://doi.org/10.1063/5.0019328} \BibitemShut {NoStop}%
\bibitem [{\citenamefont {Mahmoud}\ \emph {et~al.}(2021)\citenamefont
  {Mahmoud}, \citenamefont {Vanderveken}, \citenamefont {Adelmann},
  \citenamefont {Ciubotaru}, \citenamefont {Cotofana},\ and\ \citenamefont
  {Hamdioui}}]{amahmoud1}%
  \BibitemOpen
  \bibfield  {author} {\bibinfo {author} {\bibfnamefont {A.}~\bibnamefont
  {Mahmoud}}, \bibinfo {author} {\bibfnamefont {F.}~\bibnamefont
  {Vanderveken}}, \bibinfo {author} {\bibfnamefont {C.}~\bibnamefont
  {Adelmann}}, \bibinfo {author} {\bibfnamefont {F.}~\bibnamefont {Ciubotaru}},
  \bibinfo {author} {\bibfnamefont {S.}~\bibnamefont {Cotofana}}, \ and\
  \bibinfo {author} {\bibfnamefont {S.}~\bibnamefont {Hamdioui}},\ }\href@noop
  {} {\bibfield  {journal} {\bibinfo  {journal} {IEEE Transactions on Circuits
  and Systems I: Regular Papers}\ }\textbf {\bibinfo {volume} {68}},\ \bibinfo
  {pages} {536} (\bibinfo {year} {2021})}\BibitemShut {NoStop}%
\bibitem [{\citenamefont {Mahmoud}\ \emph
  {et~al.}(2020{\natexlab{b}})\citenamefont {Mahmoud}, \citenamefont
  {Vanderveken}, \citenamefont {Ciubotaru}, \citenamefont {Adelmann},
  \citenamefont {Cotofana},\ and\ \citenamefont {Hamdioui}}]{parallelism}%
  \BibitemOpen
  \bibfield  {author} {\bibinfo {author} {\bibfnamefont {A.}~\bibnamefont
  {Mahmoud}}, \bibinfo {author} {\bibfnamefont {F.}~\bibnamefont
  {Vanderveken}}, \bibinfo {author} {\bibfnamefont {F.}~\bibnamefont
  {Ciubotaru}}, \bibinfo {author} {\bibfnamefont {C.}~\bibnamefont {Adelmann}},
  \bibinfo {author} {\bibfnamefont {S.}~\bibnamefont {Cotofana}}, \ and\
  \bibinfo {author} {\bibfnamefont {S.}~\bibnamefont {Hamdioui}},\ }in\ \href
  {\doibase 10.23919/DATE48585.2020.9116368} {\emph {\bibinfo {booktitle} {2020
  Design, Automation Test in Europe Conference Exhibition (DATE)}}}\ (\bibinfo
  {year} {2020})\ pp.\ \bibinfo {pages} {642--645}\BibitemShut {NoStop}%
\bibitem [{\citenamefont {Mahmoud}\ \emph
  {et~al.}(2020{\natexlab{c}})\citenamefont {Mahmoud}, \citenamefont
  {Vanderveken}, \citenamefont {Adelmann}, \citenamefont {Ciubotaru},
  \citenamefont {Cotofana},\ and\ \citenamefont {Hamdioui}}]{fanout10}%
  \BibitemOpen
  \bibfield  {author} {\bibinfo {author} {\bibfnamefont {A.}~\bibnamefont
  {Mahmoud}}, \bibinfo {author} {\bibfnamefont {F.}~\bibnamefont
  {Vanderveken}}, \bibinfo {author} {\bibfnamefont {C.}~\bibnamefont
  {Adelmann}}, \bibinfo {author} {\bibfnamefont {F.}~\bibnamefont {Ciubotaru}},
  \bibinfo {author} {\bibfnamefont {S.}~\bibnamefont {Cotofana}}, \ and\
  \bibinfo {author} {\bibfnamefont {S.}~\bibnamefont {Hamdioui}},\ }\href@noop
  {} {\bibfield  {journal} {\bibinfo  {journal} {2020 IEEE Computer Society
  Annual Symposium on VLSI (ISVLSI)}\ ,\ \bibinfo {pages} {60}} (\bibinfo
  {year} {2020}{\natexlab{c}})}\BibitemShut {NoStop}%
\bibitem [{\citenamefont {Kostylev}\ \emph {et~al.}(2005)\citenamefont
  {Kostylev} \emph {et~al.}}]{logic21}%
  \BibitemOpen
  \bibfield  {author} {\bibinfo {author} {\bibfnamefont {M.~P.}\ \bibnamefont
  {Kostylev}} \emph {et~al.},\ }\href {\doibase 10.1063/1.2089147} {\bibfield
  {journal} {\bibinfo  {journal} {Appl. Phys. Lett.}\ }\textbf {\bibinfo
  {volume} {87}},\ \bibinfo {pages} {153501} (\bibinfo {year}
  {2005})}\BibitemShut {NoStop}%
\bibitem [{\citenamefont {Mahmoud}\ \emph
  {et~al.}(2020{\natexlab{d}})\citenamefont {Mahmoud}, \citenamefont
  {Vanderveken}, \citenamefont {Adelmann}, \citenamefont {Ciubotaru},
  \citenamefont {Hamdioui},\ and\ \citenamefont {Cotofana}}]{fanout}%
  \BibitemOpen
  \bibfield  {author} {\bibinfo {author} {\bibfnamefont {A.}~\bibnamefont
  {Mahmoud}}, \bibinfo {author} {\bibfnamefont {F.}~\bibnamefont
  {Vanderveken}}, \bibinfo {author} {\bibfnamefont {C.}~\bibnamefont
  {Adelmann}}, \bibinfo {author} {\bibfnamefont {F.}~\bibnamefont {Ciubotaru}},
  \bibinfo {author} {\bibfnamefont {S.}~\bibnamefont {Hamdioui}}, \ and\
  \bibinfo {author} {\bibfnamefont {S.}~\bibnamefont {Cotofana}},\ }\href@noop
  {} {\bibfield  {journal} {\bibinfo  {journal} {AIP Advances}\ }\textbf
  {\bibinfo {volume} {10}},\ \bibinfo {pages} {035119} (\bibinfo {year}
  {2020}{\natexlab{d}})}\BibitemShut {NoStop}%
\bibitem [{\citenamefont {{Mahmoud}}\ \emph {et~al.}(2020)\citenamefont
  {{Mahmoud}}, \citenamefont {{Vanderveken}}, \citenamefont {{Adelmann}},
  \citenamefont {{Ciubotaru}}, \citenamefont {{Hamdioui}},\ and\ \citenamefont
  {{Cotofana}}}]{fanout11}%
  \BibitemOpen
  \bibfield  {author} {\bibinfo {author} {\bibfnamefont {A.}~\bibnamefont
  {{Mahmoud}}}, \bibinfo {author} {\bibfnamefont {F.}~\bibnamefont
  {{Vanderveken}}}, \bibinfo {author} {\bibfnamefont {C.}~\bibnamefont
  {{Adelmann}}}, \bibinfo {author} {\bibfnamefont {F.}~\bibnamefont
  {{Ciubotaru}}}, \bibinfo {author} {\bibfnamefont {S.}~\bibnamefont
  {{Hamdioui}}}, \ and\ \bibinfo {author} {\bibfnamefont {S.}~\bibnamefont
  {{Cotofana}}},\ }in\ \href {\doibase 10.1109/ICCD50377.2020.00062} {\emph
  {\bibinfo {booktitle} {2020 IEEE 38th International Conference on Computer
  Design (ICCD)}}}\ (\bibinfo {year} {2020})\ pp.\ \bibinfo {pages}
  {332--335}\BibitemShut {NoStop}%
\bibitem [{\citenamefont {{Mahmoud}}\ \emph {et~al.}(2021)\citenamefont
  {{Mahmoud}}, \citenamefont {{Vanderveken}}, \citenamefont {{Adelmann}},
  \citenamefont {{Ciubotaru}}, \citenamefont {{Hamdioui}},\ and\ \citenamefont
  {{Cotofana}}}]{parallelism1}%
  \BibitemOpen
  \bibfield  {author} {\bibinfo {author} {\bibfnamefont {A.~N.}\ \bibnamefont
  {{Mahmoud}}}, \bibinfo {author} {\bibfnamefont {F.}~\bibnamefont
  {{Vanderveken}}}, \bibinfo {author} {\bibfnamefont {C.}~\bibnamefont
  {{Adelmann}}}, \bibinfo {author} {\bibfnamefont {F.}~\bibnamefont
  {{Ciubotaru}}}, \bibinfo {author} {\bibfnamefont {S.}~\bibnamefont
  {{Hamdioui}}}, \ and\ \bibinfo {author} {\bibfnamefont {S.}~\bibnamefont
  {{Cotofana}}},\ }\href {\doibase 10.1109/TMAG.2021.3062022} {\bibfield
  {journal} {\bibinfo  {journal} {IEEE Transactions on Magnetics}\ ,\ \bibinfo
  {pages} {1}} (\bibinfo {year} {2021})}\BibitemShut {NoStop}%
\bibitem [{\citenamefont {Mahmoud}\ \emph
  {et~al.}(2021{\natexlab{a}})\citenamefont {Mahmoud}, \citenamefont
  {Vanderveken}, \citenamefont {Adelmann}, \citenamefont {Ciubotaru},
  \citenamefont {Hamdioui},\ and\ \citenamefont {Cotofana}}]{wavepipeline}%
  \BibitemOpen
  \bibfield  {author} {\bibinfo {author} {\bibfnamefont {A.}~\bibnamefont
  {Mahmoud}}, \bibinfo {author} {\bibfnamefont {F.}~\bibnamefont
  {Vanderveken}}, \bibinfo {author} {\bibfnamefont {C.}~\bibnamefont
  {Adelmann}}, \bibinfo {author} {\bibfnamefont {F.}~\bibnamefont {Ciubotaru}},
  \bibinfo {author} {\bibfnamefont {S.}~\bibnamefont {Hamdioui}}, \ and\
  \bibinfo {author} {\bibfnamefont {S.}~\bibnamefont {Cotofana}},\ }in\ \href
  {\doibase 10.1109/ISQED51717.2021.9424264} {\emph {\bibinfo {booktitle} {2021
  22nd International Symposium on Quality Electronic Design (ISQED)}}}\
  (\bibinfo {year} {2021})\ pp.\ \bibinfo {pages} {54--59}\BibitemShut
  {NoStop}%
\bibitem [{\citenamefont {Khitun}\ \emph {et~al.}(2011)\citenamefont {Khitun}
  \emph {et~al.}}]{logic1}%
  \BibitemOpen
  \bibfield  {author} {\bibinfo {author} {\bibfnamefont {A.}~\bibnamefont
  {Khitun}} \emph {et~al.},\ }\href {\doibase 10.1063/1.3609062} {\bibfield
  {journal} {\bibinfo  {journal} {Journal of Applied Physics}\ }\textbf
  {\bibinfo {volume} {110}},\ \bibinfo {pages} {034306} (\bibinfo {year}
  {2011})}\BibitemShut {NoStop}%
\bibitem [{\citenamefont {Mahmoud}\ \emph
  {et~al.}(2021{\natexlab{b}})\citenamefont {Mahmoud}, \citenamefont
  {Vanderveken}, \citenamefont {Ciubotaru}, \citenamefont {Adelmann},
  \citenamefont {Cotofana},\ and\ \citenamefont {Hamdioui}}]{mahmoud2021spin}%
  \BibitemOpen
  \bibfield  {author} {\bibinfo {author} {\bibfnamefont {A.}~\bibnamefont
  {Mahmoud}}, \bibinfo {author} {\bibfnamefont {F.}~\bibnamefont
  {Vanderveken}}, \bibinfo {author} {\bibfnamefont {F.}~\bibnamefont
  {Ciubotaru}}, \bibinfo {author} {\bibfnamefont {C.}~\bibnamefont {Adelmann}},
  \bibinfo {author} {\bibfnamefont {S.}~\bibnamefont {Cotofana}}, \ and\
  \bibinfo {author} {\bibfnamefont {S.}~\bibnamefont {Hamdioui}},\ }in\ \href
  {\doibase 10.1109/ISCAS51556.2021.9401524} {\emph {\bibinfo {booktitle} {2021
  IEEE International Symposium on Circuits and Systems (ISCAS)}}}\ (\bibinfo
  {year} {2021})\ pp.\ \bibinfo {pages} {1--5}\BibitemShut {NoStop}%
\bibitem [{\citenamefont {Gertz}\ \emph {et~al.}(2015)\citenamefont {Gertz}
  \emph {et~al.}}]{memory3}%
  \BibitemOpen
  \bibfield  {author} {\bibinfo {author} {\bibfnamefont {F.}~\bibnamefont
  {Gertz}} \emph {et~al.},\ }\href {\doibase 10.1109/TMAG.2014.2362723}
  {\bibfield  {journal} {\bibinfo  {journal} {IEEE Trans. Magn.}\ }\textbf
  {\bibinfo {volume} {51}},\ \bibinfo {pages} {1} (\bibinfo {year}
  {2015})}\BibitemShut {NoStop}%
\bibitem [{\citenamefont {Mittal}(2016)}]{applications}%
  \BibitemOpen
  \bibfield  {author} {\bibinfo {author} {\bibfnamefont {S.}~\bibnamefont
  {Mittal}},\ }\href {\doibase 10.1145/2893356} {\bibfield  {journal} {\bibinfo
   {journal} {ACM Comput. Surv.}\ }\textbf {\bibinfo {volume} {48}} (\bibinfo
  {year} {2016}),\ 10.1145/2893356}\BibitemShut {NoStop}%
\bibitem [{\citenamefont {Kumar}\ and\ \citenamefont
  {Nath}(2017)}]{conventional_compressor4}%
  \BibitemOpen
  \bibfield  {author} {\bibinfo {author} {\bibfnamefont {M.}~\bibnamefont
  {Kumar}}\ and\ \bibinfo {author} {\bibfnamefont {J.}~\bibnamefont {Nath}},\
  }\href {\doibase 10.1088/1757-899x/225/1/012136} {\bibfield  {journal}
  {\bibinfo  {journal} {{IOP} Conference Series: Materials Science and
  Engineering}\ }\textbf {\bibinfo {volume} {225}},\ \bibinfo {pages} {012136}
  (\bibinfo {year} {2017})}\BibitemShut {NoStop}%
\bibitem [{\citenamefont {Chumak}, \citenamefont {Serga},\ and\ \citenamefont
  {Hillebrands}(2017)}]{Magnonic_crystals_for_data_processing}%
  \BibitemOpen
  \bibfield  {author} {\bibinfo {author} {\bibfnamefont {A.~V.}\ \bibnamefont
  {Chumak}}, \bibinfo {author} {\bibfnamefont {A.~A.}\ \bibnamefont {Serga}}, \
  and\ \bibinfo {author} {\bibfnamefont {B.}~\bibnamefont {Hillebrands}},\
  }\href {http://stacks.iop.org/0022-3727/50/i=24/a=244001} {\bibfield
  {journal} {\bibinfo  {journal} {Journal of Physics D: Applied Physics}\
  }\textbf {\bibinfo {volume} {50}},\ \bibinfo {pages} {244001} (\bibinfo
  {year} {2017})}\BibitemShut {NoStop}%
\bibitem [{\citenamefont {Parhami}(2009)}]{Prahami}%
  \BibitemOpen
  \bibfield  {author} {\bibinfo {author} {\bibfnamefont {B.}~\bibnamefont
  {Parhami}},\ }\href@noop {} {\emph {\bibinfo {title} {Computer arithmetic:
  Algorithms and hardware designs}}},\ Computers \& Mathematics with
  Applications\ (\bibinfo  {publisher} {Oxford University Press; 2nd edition},\
  \bibinfo {year} {2009})\BibitemShut {NoStop}%
\bibitem [{\citenamefont {{Momeni}}\ \emph {et~al.}(2015)\citenamefont
  {{Momeni}}, \citenamefont {{Han}}, \citenamefont {{Montuschi}},\ and\
  \citenamefont {{Lombardi}}}]{conventional_compressor1}%
  \BibitemOpen
  \bibfield  {author} {\bibinfo {author} {\bibfnamefont {A.}~\bibnamefont
  {{Momeni}}}, \bibinfo {author} {\bibfnamefont {J.}~\bibnamefont {{Han}}},
  \bibinfo {author} {\bibfnamefont {P.}~\bibnamefont {{Montuschi}}}, \ and\
  \bibinfo {author} {\bibfnamefont {F.}~\bibnamefont {{Lombardi}}},\ }\href
  {\doibase 10.1109/TC.2014.2308214} {\bibfield  {journal} {\bibinfo  {journal}
  {IEEE Transactions on Computers}\ }\textbf {\bibinfo {volume} {64}},\
  \bibinfo {pages} {984} (\bibinfo {year} {2015})}\BibitemShut {NoStop}%
\bibitem [{\citenamefont {{Mori}}\ \emph {et~al.}(1991)\citenamefont {{Mori}},
  \citenamefont {{Nagamatsu}}, \citenamefont {{Hirano}}, \citenamefont
  {{Tanaka}}, \citenamefont {{Noda}}, \citenamefont {{Toyoshima}},
  \citenamefont {{Hashimoto}}, \citenamefont {{Hayashida}},\ and\ \citenamefont
  {{Maeguchi}}}]{conventional_compressor2}%
  \BibitemOpen
  \bibfield  {author} {\bibinfo {author} {\bibfnamefont {J.}~\bibnamefont
  {{Mori}}}, \bibinfo {author} {\bibfnamefont {M.}~\bibnamefont {{Nagamatsu}}},
  \bibinfo {author} {\bibfnamefont {M.}~\bibnamefont {{Hirano}}}, \bibinfo
  {author} {\bibfnamefont {S.}~\bibnamefont {{Tanaka}}}, \bibinfo {author}
  {\bibfnamefont {M.}~\bibnamefont {{Noda}}}, \bibinfo {author} {\bibfnamefont
  {Y.}~\bibnamefont {{Toyoshima}}}, \bibinfo {author} {\bibfnamefont
  {K.}~\bibnamefont {{Hashimoto}}}, \bibinfo {author} {\bibfnamefont
  {H.}~\bibnamefont {{Hayashida}}}, \ and\ \bibinfo {author} {\bibfnamefont
  {K.}~\bibnamefont {{Maeguchi}}},\ }\href {\doibase 10.1109/4.75061}
  {\bibfield  {journal} {\bibinfo  {journal} {IEEE Journal of Solid-State
  Circuits}\ }\textbf {\bibinfo {volume} {26}},\ \bibinfo {pages} {600}
  (\bibinfo {year} {1991})}\BibitemShut {NoStop}%
\bibitem [{\citenamefont {Mahmoud}\ \emph
  {et~al.}(2021{\natexlab{c}})\citenamefont {Mahmoud}, \citenamefont
  {Vanderveken}, \citenamefont {Ciubotaru}, \citenamefont {Adelmann},
  \citenamefont {Hamdioui},\ and\ \citenamefont
  {Cotofana}}]{mahmoud2021spinapproximate}%
  \BibitemOpen
  \bibfield  {author} {\bibinfo {author} {\bibfnamefont {A.}~\bibnamefont
  {Mahmoud}}, \bibinfo {author} {\bibfnamefont {F.}~\bibnamefont
  {Vanderveken}}, \bibinfo {author} {\bibfnamefont {F.}~\bibnamefont
  {Ciubotaru}}, \bibinfo {author} {\bibfnamefont {C.}~\bibnamefont {Adelmann}},
  \bibinfo {author} {\bibfnamefont {S.}~\bibnamefont {Hamdioui}}, \ and\
  \bibinfo {author} {\bibfnamefont {S.}~\bibnamefont {Cotofana}},\ }\href@noop
  {} {\enquote {\bibinfo {title} {Spin wave based approximate computing},}\ }
  (\bibinfo {year} {2021}{\natexlab{c}}),\ \Eprint
  {http://arxiv.org/abs/2103.12869} {arXiv:2103.12869 [cond-mat.mes-hall]}
  \BibitemShut {NoStop}%
\bibitem [{\citenamefont {Vansteenkiste}\ \emph {et~al.}(2014)\citenamefont
  {Vansteenkiste} \emph {et~al.}}]{mumax}%
  \BibitemOpen
  \bibfield  {author} {\bibinfo {author} {\bibfnamefont {A.}~\bibnamefont
  {Vansteenkiste}} \emph {et~al.},\ }\href {\doibase 10.1063/1.4899186}
  {\bibfield  {journal} {\bibinfo  {journal} {AIP Advances}\ }\textbf {\bibinfo
  {volume} {4}},\ \bibinfo {pages} {107133} (\bibinfo {year}
  {2014})}\BibitemShut {NoStop}%
\bibitem [{\citenamefont {Devolder}\ \emph {et~al.}(2016)\citenamefont
  {Devolder} \emph {et~al.}}]{parameters}%
  \BibitemOpen
  \bibfield  {author} {\bibinfo {author} {\bibfnamefont {T.}~\bibnamefont
  {Devolder}} \emph {et~al.},\ }\href {\doibase 10.1103/PhysRevB.93.024420}
  {\bibfield  {journal} {\bibinfo  {journal} {Phys. Rev. B}\ }\textbf {\bibinfo
  {volume} {93}},\ \bibinfo {pages} {024420} (\bibinfo {year}
  {2016})}\BibitemShut {NoStop}%
\bibitem [{\citenamefont {{Manikantta Reddy}}\ \emph
  {et~al.}(2019)\citenamefont {{Manikantta Reddy}}, \citenamefont {Vasantha},
  \citenamefont {{Nithin Kumar}},\ and\ \citenamefont
  {Dwivedi}}]{approxCompCMOS1}%
  \BibitemOpen
  \bibfield  {author} {\bibinfo {author} {\bibfnamefont {K.}~\bibnamefont
  {{Manikantta Reddy}}}, \bibinfo {author} {\bibfnamefont {M.}~\bibnamefont
  {Vasantha}}, \bibinfo {author} {\bibfnamefont {Y.}~\bibnamefont {{Nithin
  Kumar}}}, \ and\ \bibinfo {author} {\bibfnamefont {D.}~\bibnamefont
  {Dwivedi}},\ }\href {\doibase https://doi.org/10.1016/j.aeue.2019.05.021}
  {\bibfield  {journal} {\bibinfo  {journal} {AEU - International Journal of
  Electronics and Communications}\ }\textbf {\bibinfo {volume} {107}},\
  \bibinfo {pages} {89} (\bibinfo {year} {2019})}\BibitemShut {NoStop}%
\bibitem [{\citenamefont {Angizi}\ \emph {et~al.}(2018)\citenamefont {Angizi},
  \citenamefont {Jiang}, \citenamefont {DeMara}, \citenamefont {Han},\ and\
  \citenamefont {Fan}}]{SPIN}%
  \BibitemOpen
  \bibfield  {author} {\bibinfo {author} {\bibfnamefont {S.}~\bibnamefont
  {Angizi}}, \bibinfo {author} {\bibfnamefont {H.}~\bibnamefont {Jiang}},
  \bibinfo {author} {\bibfnamefont {R.~F.}\ \bibnamefont {DeMara}}, \bibinfo
  {author} {\bibfnamefont {J.}~\bibnamefont {Han}}, \ and\ \bibinfo {author}
  {\bibfnamefont {D.}~\bibnamefont {Fan}},\ }\href {\doibase
  10.1109/TNANO.2018.2836918} {\bibfield  {journal} {\bibinfo  {journal} {IEEE
  Transactions on Nanotechnology}\ }\textbf {\bibinfo {volume} {17}},\ \bibinfo
  {pages} {795} (\bibinfo {year} {2018})}\BibitemShut {NoStop}%
\bibitem [{\citenamefont {Zografos}\ \emph {et~al.}(2015)\citenamefont
  {Zografos}, \citenamefont {Sorée}, \citenamefont {Vaysset}, \citenamefont
  {Cosemans}, \citenamefont {Amarù}, \citenamefont {Gaillardon}, \citenamefont
  {De~Micheli}, \citenamefont {Lauwereins}, \citenamefont {Sayan},
  \citenamefont {Raghavan}, \citenamefont {Radu},\ and\ \citenamefont
  {Thean}}]{Excitation_table_ref16}%
  \BibitemOpen
  \bibfield  {author} {\bibinfo {author} {\bibfnamefont {O.}~\bibnamefont
  {Zografos}}, \bibinfo {author} {\bibfnamefont {B.}~\bibnamefont {Sorée}},
  \bibinfo {author} {\bibfnamefont {A.}~\bibnamefont {Vaysset}}, \bibinfo
  {author} {\bibfnamefont {S.}~\bibnamefont {Cosemans}}, \bibinfo {author}
  {\bibfnamefont {L.}~\bibnamefont {Amarù}}, \bibinfo {author} {\bibfnamefont
  {P.-E.}\ \bibnamefont {Gaillardon}}, \bibinfo {author} {\bibfnamefont
  {G.}~\bibnamefont {De~Micheli}}, \bibinfo {author} {\bibfnamefont
  {R.}~\bibnamefont {Lauwereins}}, \bibinfo {author} {\bibfnamefont
  {S.}~\bibnamefont {Sayan}}, \bibinfo {author} {\bibfnamefont
  {P.}~\bibnamefont {Raghavan}}, \bibinfo {author} {\bibfnamefont {I.~P.}\
  \bibnamefont {Radu}}, \ and\ \bibinfo {author} {\bibfnamefont
  {A.}~\bibnamefont {Thean}},\ }in\ \href {\doibase 10.1109/NANO.2015.7388699}
  {\emph {\bibinfo {booktitle} {2015 IEEE 15th International Conference on
  Nanotechnology (IEEE-NANO)}}}\ (\bibinfo {year} {2015})\ pp.\ \bibinfo
  {pages} {686--689}\BibitemShut {NoStop}%
\bibitem [{\citenamefont {{Wallace}}(1992)}]{JPEG}%
  \BibitemOpen
  \bibfield  {author} {\bibinfo {author} {\bibfnamefont {G.~K.}\ \bibnamefont
  {{Wallace}}},\ }\href {\doibase 10.1109/30.125072} {\bibfield  {journal}
  {\bibinfo  {journal} {IEEE Transactions on Consumer Electronics}\ }\textbf
  {\bibinfo {volume} {38}},\ \bibinfo {pages} {xviii} (\bibinfo {year}
  {1992})}\BibitemShut {NoStop}%
\bibitem [{\citenamefont {Wang}\ \emph {et~al.}(2018)\citenamefont {Wang},
  \citenamefont {Pirro}, \citenamefont {Verba}, \citenamefont {Slavin},
  \citenamefont {Hillebrands},\ and\ \citenamefont {Chumak}}]{DC}%
  \BibitemOpen
  \bibfield  {author} {\bibinfo {author} {\bibfnamefont {Q.}~\bibnamefont
  {Wang}}, \bibinfo {author} {\bibfnamefont {P.}~\bibnamefont {Pirro}},
  \bibinfo {author} {\bibfnamefont {R.}~\bibnamefont {Verba}}, \bibinfo
  {author} {\bibfnamefont {A.}~\bibnamefont {Slavin}}, \bibinfo {author}
  {\bibfnamefont {B.}~\bibnamefont {Hillebrands}}, \ and\ \bibinfo {author}
  {\bibfnamefont {A.}~\bibnamefont {Chumak}},\ }\href@noop {} {\bibfield
  {journal} {\bibinfo  {journal} {Science Advances}\ }\textbf {\bibinfo
  {volume} {4}} (\bibinfo {year} {2018})}\BibitemShut {NoStop}%
\end{thebibliography}%

\end{document}